\documentclass[sigconf]{acmart}
\AtBeginDocument{%
  }

\copyrightyear{2026}
\acmYear{2026}
\setcopyright{cc}
\setcctype{by}
\acmConference[ICSE-SEIP '26]{2026 IEEE/ACM 48th International Conference on Software Engineering}{April 12--18, 2026}{Rio de Janeiro, Brazil}
\acmBooktitle{2026 IEEE/ACM 48th International Conference on Software Engineering (ICSE-SEIP '26), April 12--18, 2026, Rio de Janeiro, Brazil}
\acmPrice{}
\acmDOI{10.1145/3786583.3786896}
\acmISBN{979-8-4007-2426-8/2026/04}

\usepackage[utf8]{inputenc}

\PassOptionsToPackage{hyphens}{url}\usepackage{hyperref}

\usepackage{csquotes}
\usepackage{tcolorbox}
\usepackage{graphicx}
\graphicspath{{images/}}

\usepackage{multirow}

\usepackage{makecell}

\usepackage{array}

\usepackage{dblfloatfix}

\usepackage{enumitem}
\setlist[itemize]{leftmargin=20pt}

\newcommand{\answerbox}[1]{%
  \noindent\fcolorbox{customblue}{brightBackground}{%
    \begin{minipage}{\dimexpr\linewidth-2\fboxsep-2\fboxrule}
      #1
    \end{minipage}%
  }%
}

\definecolor{customblue}{RGB}{69,183,209}   
\definecolor{brightBackground}{RGB}{236,248,250} 

\begin{document}

\title[Green LLM Techniques in Action]{Green LLM Techniques in Action: How Effective Are Existing Techniques for Improving the Energy Efficiency of LLM-Based Applications in Industry?}

\author{Pelin Rabia Kuran}
\orcid{0009-0006-0166-2569}
\affiliation{
    \institution{Vrije Universiteit Amsterdam}
    \city{Amsterdam}
    \country{The Netherlands}
}
\email{pelinrkuran@gmail.com}

\author{Rumbidzai Chitakunye}
\orcid{0000-0003-0165-5650}
\affiliation{
    \institution{Vrije Universiteit Amsterdam}
    \city{Amsterdam}
    \country{The Netherlands}
}
\email{r.chitakunye@vu.nl}

\author{Vincenzo Stoico}
\orcid{0000-0002-3681-372X}
\affiliation{
    \institution{Vrije Universiteit Amsterdam}
    \city{Amsterdam}
    \country{The Netherlands}
}
\email{v.stoico@vu.nl}

\author{Ilja Heitlager}
\orcid{0000-0002-6589-3730}
\affiliation{
    \institution{Schuberg Philis}
    \city{Schiphol-Rijk}
    \country{The Netherlands}
}
\email{iheitlager@schubergphilis.com}

\author{Justus Bogner}
\orcid{0000-0001-5788-0991}
\affiliation{
    \institution{Vrije Universiteit Amsterdam}
    \city{Amsterdam}
    \country{The Netherlands}
}
\email{j.bogner@vu.nl}

\renewcommand{\shortauthors}{Kuran et al.}

\begin{abstract}
The rapid adoption of large language models (LLMs) has raised concerns about their substantial energy consumption, especially when deployed at industry scale.
While several techniques have been proposed to address this, limited empirical evidence exists regarding the effectiveness of applying them to LLM-based industry applications.
To fill this gap, we analyzed a chatbot application in an industrial context at Schuberg Philis, a Dutch IT services company.
We then selected four techniques, namely \textit{Small and Large Model Collaboration}, \textit{Prompt Optimization}, \textit{Quantization}, and \textit{Batching}, applied them to the application in eight variations, and then conducted experiments to study their impact on energy consumption, accuracy, and response time compared to the unoptimized baseline.

Our results show that several techniques, such as \textit{Prompt Optimization} and \textit{2-bit Quantization}, managed to reduce energy use significantly, sometimes by up to 90\%.
However, these techniques especially impacted accuracy negatively, to a degree that is not acceptable in practice.
The only technique that achieved significant and strong energy reductions without harming the other qualities substantially was \textit{Small and Large Model Collaboration} via Nvidia's Prompt Task and Complexity Classifier (NPCC) with prompt complexity thresholds.
This highlights that reducing the energy consumption of LLM-based applications is not difficult in practice.
However, improving their \textit{energy efficiency}, i.e., reducing energy use without harming other qualities, remains challenging.
Our study provides practical insights to move towards this goal.
\end{abstract}




\settopmatter{printfolios=true}

\maketitle

\section{Introduction}\label{s:intro}
Large language models (LLMs) have become increasingly integral to our daily lives, facilitating a range of tasks from basic automation to complex problem-solving~\cite{chang_survey_2024}.
However, this development comes with significant downsides, particularly in terms of their energy consumption~\cite{argerichQuantiz}.
Besides the energy consumption concerns, the widespread public use of online chatbot applications such as ChatGPT and Gemini may raise data privacy and security concerns for some organizations~\cite{llm_security_cool_paper}.
This may encourage these organizations to develop internal chatbot systems that rely on publicly available LLMs within more controlled environments.

An example of such an internal chatbot platform is \textit{ChatSBP}, the LLM-based application used as a case study object in this paper.
ChatSBP is developed by Schuberg Philis (SBP), an IT services company based in the Netherlands focusing on building mission-critical software.\footnote{https://schubergphilis.com}
It integrates state-of-the-art LLMs to support a range of language-based tasks, from basic text correction to complex problem solving.
While these capabilities offer substantial utility, the increasing usage of ChatSBP not only starts to add substantial costs but also fuels existing concerns regarding the environmental impact of SBP and similar LLM-using companies.

As awareness of energy consumption and carbon emissions grows, optimizing LLM inference for sustainability has become a key concern~\cite{adamska2025greenprompting}.
In response, reusable green techniques have started to emerge~\cite{jarvenpaa_synthesis_2024}, such as model quantization~\cite{ma2025onebit}, batching~\cite{debated_batching_image}, prompt optimization~\cite{agarwal2024promptwizard}, and large and small model collaborations~\cite{minions}.
However, especially for real-world LLM-based applications, the applicability, effectiveness, and trade-offs of these techniques, e.g., regarding response time and accuracy, remain insufficiently examined~\cite{argerichQuantiz,poddar2025measure_nlp}.

In this case study, we therefore investigated the practical effectiveness of selected energy optimization techniques when applied to an LLM-based industrial chatbot application (ChatSBP).
Specifically, we analyzed how effective four existing energy optimization techniques in several configurations are in reducing energy consumption and what potential trade-offs exist regarding accuracy and latency, two crucial quality concerns in most industrial settings.
We accomplished this by individually applying the techniques to ChatSBP to create several variants and then ran controlled experiments on each variant and the unoptimized baseline application in SBP's industrial experiment environment.

\section{Related Work}
\label{s:related_work}
LLMs have shown growing capabilities in various domains~\cite{chang_survey_2024}, most notably in software engineering~\cite{hou_large_2024}.
However, their continuous growth in size also contributes to increased use of computational resources and energy.
Research has shown that larger models also consume more energy~\cite{Geens2024, argerichQuantiz, stojkovic2025dynamollm, Khan2025, Rubei2025, Kaushik2025}.
In an effort to promote green artificial intelligence (AI), i.e., AI that is accurate, environmentally friendly, and inclusive~\cite{Schwartz2020}, researchers have explored different optimization techniques to find favorable trade-offs between energy consumption and accuracy~\cite{stojkovic2025dynamollm, Geens2024, argerichQuantiz, Rubei2025}.
Most research has focused on the training phase, although previous studies have shown that more energy is consumed during inference at scale~\cite{argerichQuantiz, Chien2023}.
We describe selected studies looking at individual optimization techniques for LLM inference below.

\textit{Small language models} (SLMs) rely on reduced size compared to LLMs, as their parameter count ranges only from millions to a few billions.
Their size makes them ideal for execution on resource-constrained devices, but they may struggle to match the accuracy of LLMs~\cite{wang2024comprehensivesurveysmalllanguage}. 
Literature focuses on investigating how SLMs can be used while keeping acceptable levels of accuracy and maintaining their lower resource usage.
For example, \citet{kavathekar2025small} studied the ability of SLMs to generate function calls using zero- and few-shot prompting as well as fine-tuning.
They evaluated the accuracy, robustness to prompt injection, and resource usage of multiple SLMs.
Fine-tuned models outperformed zero- and few-shot counterparts significantly.
For example, Phi-2 achieved 62.4\% accuracy server-side and maintained reasonable latency on edge devices ($\sim$140s).
Despite the analysis of SLMs resource usage, the work misses a discussion of their energy usage, which we add in our study.
In a similar study, \citet{duran2024energy} executed a controlled experiment on SLMs with a total of 600 inference requests across 12 different models and 5 configurations, such as the baseline ⟨TORCH, CUDA⟩ setup.
They evaluated energy usage, execution time, and resource usage, namely CPU, memory, and GPU usage.
The results showed that the chosen configuration has significant impact on energy and execution time.
Specifically, the configurations using CUDA consistently reduced energy use and inference time.
While their study provides valuable insights about the green deployment of SLMs, their testbed is largely a controlled experimental setup rather than a real-world application.
In contrast, our study takes a system perspective and integrates SLMs into an industrial application, ChatSBP, deployed at Schuberg Philis.

Another promising technique is the \textit{batching} of inference requests.
For example, \citet{argerichQuantiz} showed that batching can maximize resource utilization and reduce energy consumption.
Their study evaluated the impact of model architectures and sizes as well as batch sizes and quantization on the energy usage of 15 LLMs.
The results showed that maximizing batch size reduced the energy usage of all models by up to 20 times, while it could increase GPU usage to its full capacity.
The strength of the energy reduction depended on the LLM architecture.
Similarly, \citet{Walkowiak2025} implemented continuous batching in text generation to study the effect of batch sizes on energy consumption. 
Results indicate that batch sizes from 1 to 150 reduced energy use, although they negatively affected response time.

Another proposed technique is \textit{prompt-level optimization}, which entails crafting more precise and context-aware prompts, making it possible to obtain higher quality responses while saving other resources like time or computation~\cite{Ggaliwango2024, agarwal2024promptwizard, adamska2025greenprompting}.
As a concrete example, \citet{agarwal2024promptwizard} introduced the PromptWizard framework to improve task performance. 
PromptWizard automates prompt optimization by using feedback-driven critique to refine prompts and in-context examples.
Trying a different angle, \citet{Rubei2025} investigated to what extent custom tags in prompts reduce energy consumption during inference. 
They compared three different prompt engineering techniques implemented with and without custom tags.
Their results indicate that the use of custom tags in all three prompt engineering techniques reduced LLM energy consumption.
Both studies were limited to prompt engineering with a focus on optimizing and improving existing techniques.

\textit{Quantization} reduces model size and computational demands by representing the weights and activations with fewer bits~\cite{argerichQuantiz}.
For example, \citet{Geens2024} implemented weight-only quantization by comparing the two techniques W4A16 and W1A32.
Their results indicate that weight-only quantization can reduce both energy consumption and latency.
Similarly, \citet{Husom2025} investigated the effect of quantization on energy consumption, inference speed, and trade-offs between accuracy and energy use across different quantization levels during LLM edge inference.
Their findings showed that quantization had notable positive impacts on both energy consumption and inference latency, with heavy dependence on the concrete task.
Despite the empirical evidence of reduced energy consumption in both studies, the studies were implemented on an experimental basis and focused only on quantization.
  
In summary, the studies by \citet{Khan2025} and \citet{argerichQuantiz} exhibit some similarities with our research by considering more than one optimization technique.
\citet{Khan2025} used post-quantization and local inference techniques to reduce the energy consumption of an LLM by 45\%.
\citet{argerichQuantiz} measured how the size, architecture, batch size, and weight quantization of the model affect the energy consumption and latency of LLM text generation.
Each of these studies focused on one task, thus lacking task variety when evaluating their optimization techniques.
Additionally, the experiments conducted in both studies do not focus on the trade-offs of reduced energy consumption and other quality attributes.
Finally, neither study takes a pronounced system perspective and applies the techniques to real-world applications.
To fill the mentioned gaps, we apply a variety of optimization techniques to an existing industry application to provide detailed insights into their practicality in the real world and to investigate the trade-offs between energy consumption and other quality attributes, specifically response time and accuracy.

\section{Study Design}\label{s:design}
To start closing the described research gap, we formed an academia-industry collaboration between VU Amsterdam and Schuberg Philis (SBP)\footnote{\url{https://schubergphilis.com}}, a Dutch software \& IT services company with about 450 employees.
SBP focuses on mission-critical IT systems and has a strong commitment to digital sustainability.\footnote{\url{https://schubergphilis.com/our-esg-sustainability-commitment}}
Identifying effective techniques to improve the energy efficiency of their internal applications is therefore not only an attractive means to save costs but a matter of principle.

Together, we designed and executed a \textit{case study}~\cite{runeson_guidelines_2009} using ChatSBP as the central object of our investigation.
We first studied the architecture and technologies of ChatSBP and then identified, discussed, and selected a set of green techniques to improve the energy efficiency of ChatSBP.
Finally, we conducted \textit{controlled experimentation}~\cite{Wohlin2024} to collect quantitative evidence on the different versions of the application in the Leaplab at SBP, an industrial experiment environment with hardware-based energy measurement.

Our study was guided by the following research questions:

\begin{enumerate}
    \item[\textbf{{RQ}1:}] How effective are proposed optimization techniques in reducing the energy consumption of industrial LLM-based applications?
    \item[\textbf{{RQ}2:}] What are the trade-offs of the selected techniques with other quality attributes?
\end{enumerate}

With RQ1, we wanted to investigate the real-world effectiveness of the selected green techniques by measuring energy consumption before and after implementing them.
The objective is to assess whether these techniques lead to tangible sustainability benefits when evaluated under realistic application requirements.
With RQ2, we explored the trade-offs of the applied energy optimization techniques in terms of two important quality attributes: response time and accuracy.
Both of these are very important for chatbot applications in industry, i.e., even if a technique is effective in reducing energy consumption, it may not be suitable for many application scenarios if it hurts accuracy and response time substantially.
Lastly, we also studied how the techniques impacted the LLM output token counts, as this attribute is related to costs but is also sometimes used as a predictor for response time and energy consumption if direct measurements are not possible.

\subsection{Case Study Object: ChatSBP}
\begin{figure}[h]
    \centering
    \includegraphics[width=0.75\linewidth]{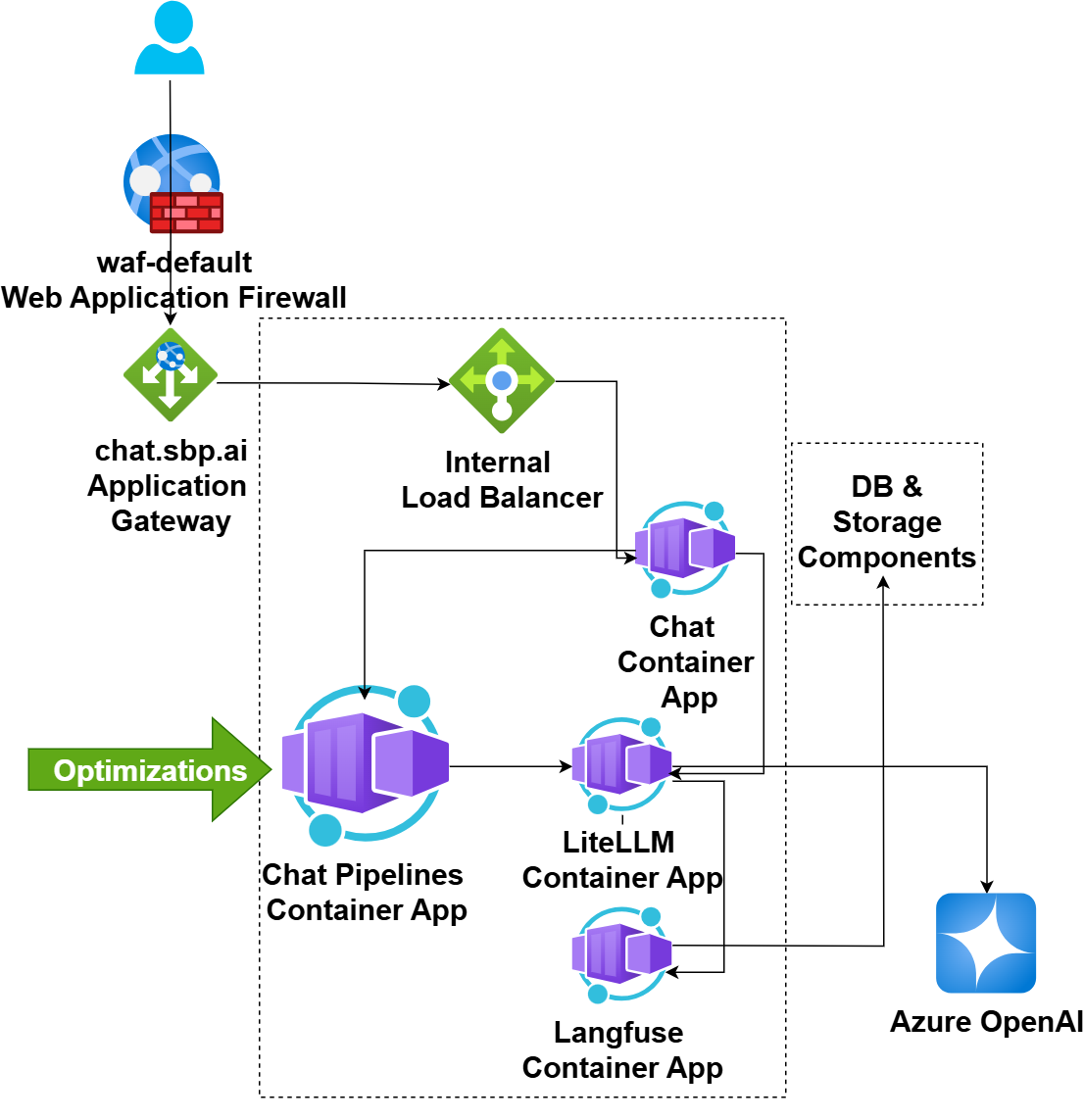}
    \caption{ChatSBP Software Architecture and the Target of Our Optimizations}
    \label{fig:SystemArchitecture}
\end{figure}

ChatSBP is the internal chatbot application used at SBP, which provides a customized environment for using LLMs, including fine-tuned models.
Through a web user interface, SBP employees may select any of the LLMs that the company currently offers for assistance with coding for software development and any other tasks.\footnote{\url{https://schubergphilis.com/stories/ai-and-automation}}
The software architecture of ChatSBP consists of a dockerized web application built on top of OpenWebUI.\footnote{\url{https://github.com/open-webui/open-webui}}
The application usually runs on cloud infrastructure of Microsoft Azure \footnote{\url{https://azure.microsoft.com}} behind a firewall and a load balancer.
The core software architecture comprises four components relevant to LLM inference (see Fig.~\ref{fig:SystemArchitecture}):

\begin{itemize}
    \item \textbf{Chat Container App}: Frontend component where users submit their prompts.
    \item \textbf{Chat Pipelines Container App}: Manages application logic and workflows. Each pipeline processes user inputs and generates outputs in a predefined order. Energy optimization techniques are incorporated into the software through this container, with each technique implemented as a new pipeline
    \item \textbf{Database}: Stores chat history.
    \item \textbf{Azure OpenAI}: Serves as the LLM inference endpoint, processing user prompts and generating responses.
\end{itemize}

In the production instance, LLM inference is managed externally by the Azure OpenAI API.
Therefore, we could not directly measure energy consumption within this virtualized cloud environment.
To obtain reliable energy measurements, we deployed all versions of the application in the local experiment environment of SBP, the Leaplab, which enables direct power monitoring through dedicated hardware instrumentation.
We applied all optimizations to the Chat Pipelines Container App, which was the only modified component.
All other components remained in their original state.

\begin{table*}[ht]
    \centering
    \small
    \caption{Selected Energy Optimization Techniques}
    \label{tab:selected-techniques}
    \begin{tabular}{lp{0.47\linewidth}l}
        \textbf{Technique} & \textbf{Description and Rationale} & \textbf{Implementation Variations} \\ 
        \hline
        \hline
        T1: Small and Large LM Collaboration~\cite{minions} & Route suitably simple prompts to a smaller, local LM instead of to the default energy-hungry LLM & \makecell[tl]{T1AA: NPCC 0.6\\T1AB: NPCC 0.3\\T1B: Minion}\\
        T2: Prompt Optimization~\cite{agarwal2024promptwizard}  & Make prompts clearer and more concise to reduce the token count and therefore the latency and ideally the energy use & PromptWizard\\
        T3: Quantization~\cite{poddar2025measure_nlp} & Reduce the precision of how model weights are represented to a smaller number of bits, which is computationally less demanding and therefore ideally requires less energy & \makecell[tl]{T3C: 2-Bit Model (Phi-4 14B)\\T3D: 4-Bit Model (Phi-4 14B)\\T3E: 8-Bit Model (Phi-4-mini 3.8B)}\\
        T4: Batching~\cite{argerichQuantiz}  &  Sending multiple, batched queries to the model instead of just a single one enables better parallelization and optimizes CPU/GPU utilization, which reduces latency per token and therefore ideally also energy consumption & T4: 2 queries per batch\\
        \hline
        \hline
    \end{tabular}
\end{table*}

\subsection{Energy Optimization Techniques}\label{s:implementation}
Since there are many available green techniques for ML-enabled systems, we followed a three-step approach to select a suitable number of techniques for our study.
First, we used the collection of techniques described in~\citet{jarvenpaa_synthesis_2024} and more recent LLM-focused publications (see the replication package for the complete list of papers) to identify potential energy optimization techniques.
Second, we analyzed the software architecture of ChatSBP to understand technical constraints and integration points. Finally, we formulated and applied a set of inclusion and exclusion criteria to guide our selection based on practical deployment requirements.

\textbf{Inclusion Criteria:} (i) the technique is compatible with the technology stack of ChatSBP, (ii) the technique can be implemented with reasonable effort, e.g., the implementation can be customized for another platform, and (iii) the technique is promising to reduce the energy consumption of the chat functionality, not of other system components.

\textbf{Exclusion Criteria:} (i) the technique introduces unsuitably high computational complexity for a real-time chatbot application, and (ii) the technique has no reusable implementation, which prevents replicability and may create bias through a custom implementation.

One important argument from SBP was that token counts and the number of API calls likely directly impact energy consumption but also cloud provider costs~\cite{azure-openai-pricing,azure-phi3-pricing}.
Techniques that reduce token counts, the reliance on larger models, or redundant requests were therefore prioritized.
SBP colleagues also highlighted potential inefficiencies that increase costs due to unnecessary LLM calls, such as vague or open-ended prompts (\enquote{hello}, \enquote{help}) or polite tokens (\enquote{please}, \enquote{thanks}).
After thorough discussion, we finally selected four energy optimization techniques (see Table~\ref{tab:selected-techniques}).
The complete list of considered techniques can be found in our replication package.\footnote{\label{fn:replication-pckg}\url{https://doi.org/10.5281/zenodo.16462531}}

First, \textit{Small and Large Language Model Collaboration} (T1) emerged as an attractive technique for ChatSBP.
Smaller models can handle simple queries locally, reducing unnecessary invocations of more energy-hungry LLMs.
Moreover, a decision layer that routes the query to either the small or large model can be easily integrated into the Chat Pipelines Container.
We implemented three different T1 variants: two implementations using Nvidia's Prompt Task and Complexity Classifier (NPCC)~\cite{nvidia_prompt_task_complexity_classifier} with different complexity thresholds (0.3 and 0.6) and one using the Minion framework~\cite{minions} for language model collaboration with default parameters.

Second, \textit{Prompt Optimization} (T2) improves accuracy, but there are also indications that it could improve energy efficiency~\cite{adamska2025greenprompting,apsan2025generating}.
Especially for chatbot applications, automatic prompt optimization is promising to minimize inefficient queries, API calls, and token usage~\cite{agarwal2024promptwizard}.
To implement this technique, we used the popular PromptWizard library\footnote{\url{https://github.com/microsoft/PromptWizard}} from Microsoft with default parameters.

Third, \textit{Quantization} (T3) represents the LLM weights and activations with lower-precision data types~\cite{argerichQuantiz}, usually starting from 16 bits and going down to as low as 1 bit~\cite{ma2024era1bitllmslarge}.
This increases their computational efficiency but may also reduce accuracy.
Nonetheless, quantized models often retain sufficient accuracy to avoid larger models.
Energy consumption is also reduced, but some studies argue that quantization below 4 bits does not lead to significant reductions~\cite{argerichQuantiz,poddar2025measure_nlp}.
However, recent research by \citet{NEURIPS2023_0df38cd1} implies otherwise for 2-bit quantization.
We therefore included three T3 variants: a 2-bit, 4-bit, and 8-bit LLM.

Lastly, \textit{Batching} (T4) executes LLM inference for several prompts in parallel~\cite{poddar2025measure_nlp}, which leads to better resource utilization and ideally more energy efficiency.
However, the concrete impact of larger batch sizes on energy consumption is still debated~\cite{argerichQuantiz,debated_batching_image}.
Since it is conveniently supported by most LLMs and easy to implement in the Chat Pipelines Container, we therefore included batching with a batch size of 2 queries.

\subsection{Experimental Materials}
\subsubsection{Language Models}
To implement all ChatSBP variants, we needed one small LM, one large LM, and the three quantized LMs.
We decided to select all five models from the same model family to allow a fairer comparison between the treatments.
For the small model, we chose Phi4-mini 3.8B, which outperforms larger models like Llama 3.1~\cite{microsoft2025phi4minitechnicalrepo} and offers a good compromise between size, energy use, and accuracy in resource-limited settings.
For the default large LM, we selected Phi4, a GPT-like decoder-only model.
Despite only having 14 billion parameters, Phi4 showed strong benchmark performance, surpassing some state-of-the-art models with up to 70 billion parameters~\cite{abdin2024phi4technicalreport}.
For the quantized models (2-bit, 4-bit, and 8-bit), we used Unsloth-AI\footnote{\url{https://unsloth.ai}} as a popular open provider for quantized models. 
The selected models were Phi-4-GGUF\footnote{\url{https://huggingface.co/unsloth/phi-4-GGUF}} with the formats Q2\_K and Q4\_K\_M and Phi-4-mini-instruct-GGUF\footnote{\url{https://huggingface.co/unsloth/Phi-4-mini-instruct-GGUF}} with the format Q8\_0.
We included the Phi-4-mini (base\_s) model in this study due to hardware limitations, as we could not run the 8-bit quantized version of the full Phi-4 model on our system.
Additionally, we wanted to observe the impact of increased quantization and gain insights by comparing the small and large models in terms of the defined quality attributes.
For all models, temperature was set to 0 to have deterministic responses for the experiment~\cite{renze-2024-effect}.

\subsubsection{Benchmark Datasets}
Since SBP did not have a custom benchmark for queries relevant to their work, we instead selected two different datasets for the experiment.
The goal was to include tasks that are very different from each other to understand if technique effectiveness was dependent on specific task types, but also to use tasks with some structural similarity to queries used at SBP, despite not being from the same domain.
For reasoning tasks, we chose \textit{GSM8K} (Grade School Math 8K)~\cite{gsm8k}, which includes solving arithmetic problems or word problems that combine logical reasoning with mathematical operations.
These difficult exercises test the ability of a model to reason about and solve problems requiring mathematical operations, logical thinking, and multi-step inference.
Additionally, for Q\&A-oriented prompts, we selected \textit{MMLU} (Massive Multitask Language Understanding)~\cite{mmlu}, one of the most commonly used datasets to evaluate a model's ability to generate accurate and relevant answers to multiple-choice questions.

\begin{figure}[h]
    \centering
    \includegraphics[width=0.7\linewidth]{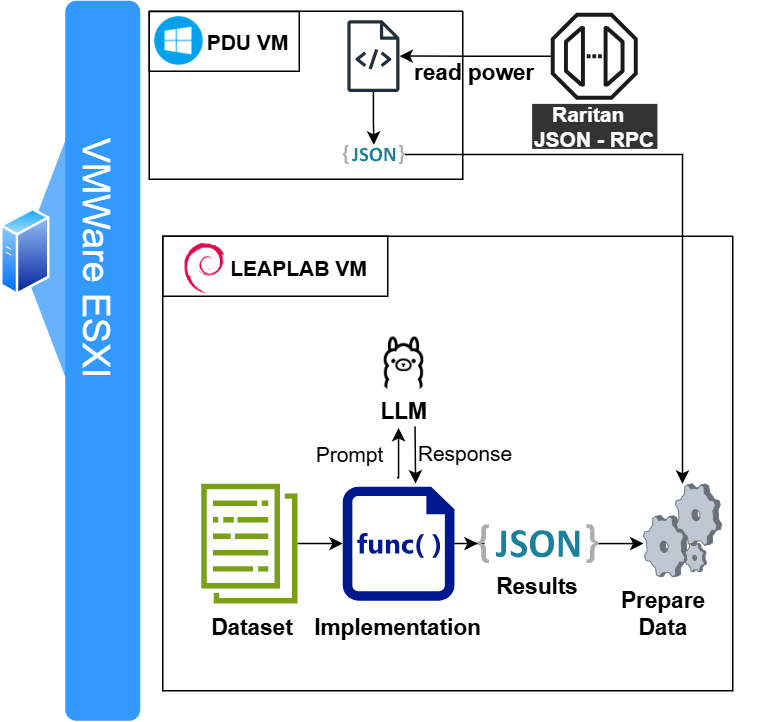}
    \caption{Experiment Infrastructure (Leaplab)}
    \label{fig:expArch}
\end{figure}

\subsubsection{Experiment Infrastructure}
\label{subsection:leaplab}
To have full control over the experiment environment, we used the Leaplab at SBP instead of the Azure cloud environment of the production instance (see Fig.~\ref{fig:expArch}).
It uses a VMware ESXi hypervisor \footnote{\url{https://www.vmware.com/products/cloud-infrastructure/vsphere}} powered by an Intel(R) Xeon(R) Gold 6154 CPU @ 3.00 GHz server.
Two virtual machines were used: an experiment VM (Leaplab VM) and a Power Distribution Unit VM (PDU-VM).
The experiment VM ran as the only ESXi workload and used a dedicated GPU.
We used a PDU outlet for power reading, which could be easily converted into energy consumption data without 3rd-party tools.
Each VM used separate PDU outlets to avoid measurement interference between them.
Energy consumption data was accessed via the Raritan JSON-RPC API.
The PDU-VM has 4 GB RAM and runs on Windows Server 2019 Standard (64-bit).
The Leaplab VM runs on Debian 12 with 32 GB of RAM and is equipped with an Nvidia A2 Tensor Core GPU with 16 GB of RAM.
Lastly, we used the Ollama framework\footnote{\url{https://ollama.com}} v0.6.6 to deploy the LLMs for the experiments.

\subsection{Experiment Design and Execution}
As \textit{independent variables}, our experiment included the optimization techniques with all implementation variations (8 in total as displayed in Table~\ref{tab:selected-techniques}, plus a not optimized baseline) and the two different benchmark datasets (GSM8k and MMLU).
Our \textit{dependent variables} consisted of the energy consumption of the system (Wh), accuracy, response time (ms), and output token count.
We measured the energy consumption of the system through Raritan / Server Technology Xerus™ PDU interface via a JSON-RPC API.\footnote{\url{https://help.raritan.com/json-rpc/4.3.0/index.html}}
Accuracy is measured using a simple string matching algorithm, as the questions from the benchmark datasets have a single correct answer.
Response time as an indicator for energy consumption in black box models is debated.
Some argue that longer response times in LLM-based systems always mean higher energy usage~\cite{adamska2025greenprompting,poddar2025measure_nlp}, while others disagree~\cite{argerichQuantiz}.
Gathering both metrics in a controlled environment provides insight into their relationship, but response time is also a valuable metric on its own: techniques that increase response time too much may not be usable in practice, particularly for chatbots.
Lastly, output token count as more of an explanatory variable is operationalized via the HuggingFace \texttt{AutoTokenizer} class.\footnote{\url{https://huggingface.co/docs/transformers/main/en/model_doc/auto\#transformers.AutoTokenizer}}
Collecting output token counts increases the granularity of our performance analysis and helps to get additional insights when trying to interpret differences between techniques.
 
For evaluating the effect of optimization techniques on energy consumption (RQ1), we formulated the following hypothesis pairs, where $e$ denotes energy consumption, $x$ a specific technique, and $\mu^e_{x}$ the median energy consumption for technique $x$:
\[
H^x_0: \mu^e_{x} \geq \mu^e_{\text{baseline}}
\]
\[
H^x_1: \mu^e_{x} < \mu^e_{\text{baseline}}
\]
For each evaluated technique, the null hypothesis was that the median energy consumption of the system variant that implemented the technique was greater than or equal to the energy consumption of the baseline system.
We used the median instead of the mean because we expected our groups to be non-normally distributed.
The alternative hypothesis then was that median energy consumption of the technique was less than the baseline.
For RQ2, we relied on visual inspection and descriptive statistics.
Thus, no formal hypotheses were formulated.

With both independent variables as factors, we used a full factorial design~\cite{Wohlin2024} to establish the experiment configurations.
This led to 8 technique implementations plus a baseline, i.e., 9 system variations $\times$ 2 datasets = 18 experiment configurations.
Before executing the experiment, non-essential background processes were terminated to minimize potential measurement impact on dependent variables.
For additional measurement reliability, each of the 18 experiment configurations was iterated 5 times, leading to a total of 90 experiment trials.
To allow the experiment environment to return closer to the initial temperature, we also introduced a cool-down period of five minutes between each trial.
Additionally, the scheduling of treatment trials was randomized to avoid the potential short-lived confounders that disproportionately affected a single treatment.
Finishing all 90 trials took roughly 24 hours.

\subsection{Data Analysis}
For statistical evaluation, we first checked whether the data followed a normal distribution using Q-Q plots for graphical evaluation~\cite{q-qtest} and the Shapiro-Wilk test for statistical evaluation~\cite{shapiro}.
A deviation from the expected diagonal in the Q-Q plots and p-values below 0.05 in the Shapiro-Wilk test indicated non-normality in the results.
We also used boxplots and descriptive statistics to understand the differences between treatments better.
We observed a non-normal distribution in our groups and therefore used a non-parametric method for hypothesis testing, namely the Mann-Whitney U test~\cite{Mann-Whitney}, to compare energy consumption between baseline and treatment groups, separately for each dataset.
A one-sided test was chosen to detect if a technique led to a significant decrease.
Before hypothesis testing, we applied the interquartile range (IQR) method with the default 1.5 factor to filter out potential outliers from the energy consumption data~\cite{vinutha2018iqr}.
This mitigated the influence of potential extreme values occurring due to unwanted side effects in the experiment environment, which may skew the results, making visualizations more difficult to interpret.
Since we conducted 18 hypothesis tests in total, we applied the Holm-Bonferroni correction to adjust p-values and control the family-wise error rate~\cite{Holm-Bonferroni}.
Lastly, we calculated effect sizes via the non-parametric Cliff’s $\delta$ to assess the strength of significant effects~\cite{cliffs'd}.

\begin{figure*}
    \centering
    \includegraphics[width=0.8\linewidth]{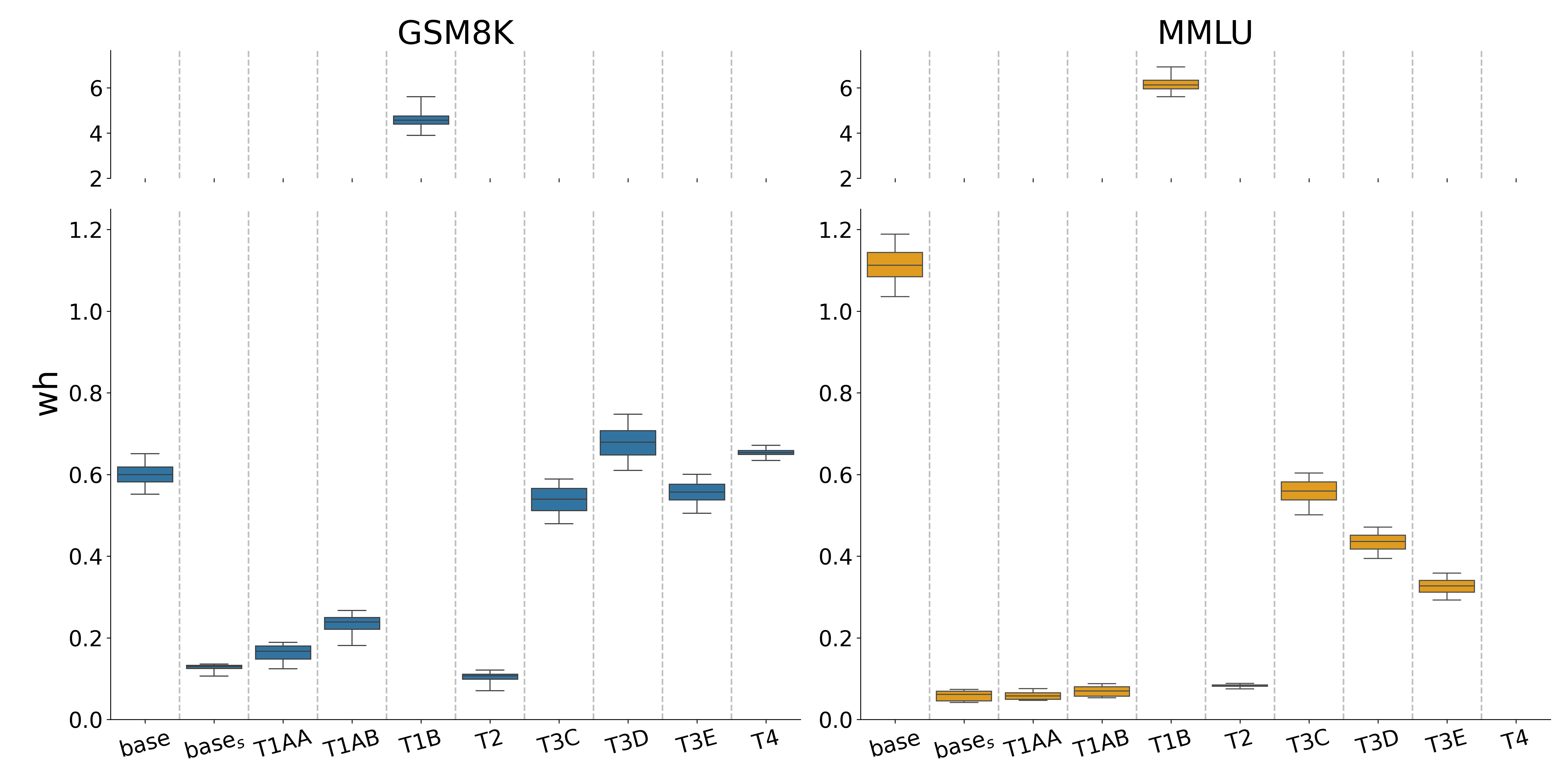}
    \caption{Energy Consumption per Prompt in Wh for Each Treatment {\normalfont\small(base: \textit{Phi4}, base$_s$: \textit{Phi4-mini} (only compared to T3E), T1AA: \textit{NPCC 0.6}, T1AB: \textit{NPCC 0.3}, T1B: \textit{Minion}, T2: \textit{PromptWizard}, T3C: \textit{2-bit Quant.}, T3D: \textit{4-bit Quant.}, T3E: \textit{8-bit Quant.}, T4: \textit{Batching with size 2})}}
    \label{fig:energy_cons}
\end{figure*}

\section{Results}\label{s:evaluation}
In this section, we present the study results structured by the research questions.
For transparency and replicability, we share all artifacts, datasets, codebases, and experiment data on Zenodo.\textsuperscript{\ref{fn:replication-pckg}}

\begin{table}[ht]
    \caption{Descriptive Statistics for Energy Consumption (Wh) per Prompt and Experiment Configuration {\normalfont\small(baseline$_s$: Phi4-mini, only compared to T3E; T4 MMLU missing due to hardware constraints)}}
    \label{tab:descriptive_energy}
    \centering
    \small
    \begin{tabular}{llrrrrr}
        \textbf{Dataset} & \textbf{Technique} & \textbf{Mean} & \textbf{STD} & \textbf{Min} & \textbf{Median} & \textbf{Max} \\
        \hline
        \hline
        \multirow{10}{*}{\textbf{GSM8k}}
                         & T1AA               & 0.16        & 0.02       & 0.11      & 0.16         & 0.20       \\  
                         & T1AB               & 0.23        & 0.02       & 0.16       & 0.23         & 0.30      \\  
                         & T1B                & 4.56        & 0.26      & 3.75       & 4.55         & 5.40       \\  
                         & T2                 & 0.10        & 0.003       & 0.07       & 0.10          & 0.13       \\  
                         & T3C                & 0.53        & 0.04       & 0.44       & 0.53          & 0.65       \\  
                         & T3D                & 0.67       & 0.04       & 0.56       & 0.67          & 0.80       \\  
                         & T3E                & 0.56        & 0.03       & 0.49       & 0.55          & 0.63       \\  
                         & T4            & 0.65        & 0.008       & 0.63      & 0.65          & 0.67       \\  
                          & baseline      & 0.59  & 0.02  & 0.51  & 0.59  & 0.68 \\ 
                         & baseline$_s$  & 0.13  & 0.005  & 0.11  & 0.13  & 0.14 \\
        \hline
        \multirow{10}{*}{\textbf{MMLU}}
                         & T1AA               & 0.05       & 0.01       & 0.04       & 0.05         & 0.08     \\  
                         & T1AB               & 0.07       & 0.01       & 0.05       & 0.06          & 0.11       \\  
                         & T1B                & 6.13       & 0.27      & 5.28      & 6.12          & 7.00      \\  
                         & T2                 & 0.08       & 0.003       & 0.07       & 0.08          & 0.09       \\  
                         & T3C                & 0.55        & 0.03       & 0.46       & 0.55          & 0.65       \\  
                         & T3D                & 0.43        & 0.02       & 0.35       & 0.43          & 0.51       \\  
                         & T3E                & 0.32        & 0.02       & 0.26       & 0.32          & 0.38     \\
                          & T4 & -- & -- & -- & -- & --\\
                          & baseline     & 1.11  & 0.05  & 0.98  & 1.11  & 1.25 \\ 
                        & baseline$_s$  & 0.06  & 0.01  & 0.04  & 0.06  & 0.08 \\
        \hline
        \hline
    \end{tabular}
\end{table}

\subsection{Reducing Energy Consumption (RQ1)}
With RQ1, we wanted to study how effective the eight technique variations were in reducing energy consumption compared to the unoptimized baseline.
We show the energy consumption distribution for the baseline and each selected technique separately for each dataset in Fig.~\ref{fig:energy_cons}.
Additionally, we present the respective descriptive statistics in Table~\ref{tab:descriptive_energy}.
Averaged per prompt, mean energy consumption ranged roughly from 0.1 Wh to 6 Wh, which suggests substantial differences between experiment configurations.
Compared to the baseline, several techniques achieved a visible energy reduction in both benchmarks, namely the two NPCC configurations (T1AA and T1AB), PromptWizard (T2), and 2-bit quantization (T3C).
Other techniques like 4-bit quantization (T3D) were only successful with the MMLU dataset.
Lastly, Minion (T1B), batching (T4), and 8-bit quantization (T3E), which was compared to the small base$_s$, were not able to lower energy consumption at all, with T1B even leading to a substantial increase.
For batching (T4) with the MMLU dataset, our hardware setup could not handle the context length of two combined prompts (see Section~\ref{subsection:leaplab}), which is why Table~\ref{tab:descriptive_energy} shows dashes instead of statistics in that row.

We used the Mann-Whitney U test to confirm if these observed differences to the baseline were statistically significant.
Table~\ref{tab:cliffs_energy} presents the test results with Holm-Bonferroni-corrected p-values, effect sizes, and percentage differences of the medians.
The most impactful energy reductions were observed for T1AA, T1AB, T2, and T3C, which were statistically significant in both datasets and showed large effect sizes (Cliff's $\delta$ between 0.77 and 1.0), indicating substantial energy savings compared to the baseline.
Especially techniques like PromptWizard (T2) and NPCC configurations (T1AA and T1AB) achieved reductions of at least 60\% on the reasoning dataset (GSM8k) and at least 92\% on the Q\&A dataset (MMLU).

In terms of individual techniques, the impact of the \textit{Small and Large LM Collaboration} technique (T1) was strongly dependent on the concrete implementation variation.
Minion (T1B) substantially increased energy consumption in both datasets by more than 450\%, which was mostly due to several message exchanges between the small and the large model and several cases of excessive token generation by the small model.
However, the NPCC variations were among the most effective reduction techniques.
Using a prompt complexity threshold of 0.6 (T1AA) resulted in slightly lower energy consumption compared to the 0.3 threshold of T1AB, e.g., median energy consumptions within GSM8k were 0.16 Wh vs. 0.23 Wh.
A higher threshold meant that the more energy-hungry LLM was called less often.
For the reasoning dataset (GSM8k), the smaller model was used 72\% of the time with T1AB, while this increased to 86\% with T1AA.
For the Q\&A dataset (MMLU), the smaller model was used approximately 98\% of the time in both T1AA and T1AB, which explains why the median energy consumption was much closer this time (0.05 Wh vs. 0.06 Wh), but also why the NPCC techniques were even more effective within MMLU compared to GSM8k.

The \textit{Prompt Optimization} technique (T2) via PromptWizard achieved the highest average energy reductions across both datasets, with 82\% for GSM8k (the highest) and 92.5\% for MMLU (the 3rd highest, but the first three were very close together with 92.5\% to 93.7\%).
It also had the lowest median (0.10 Wh) and maximum (0.13 Wh) energy consumption per prompt for the more complex GSM8k dataset, which further demonstrates its effectiveness.

\begin{table}
    \centering
    \small
    \caption{Energy Consumption Baseline Comparisons via Mann-Whitney U Tests {\normalfont\small(Holm-Bonferroni-corrected p-values, significant effect sizes via Cliff’s $\delta$, median percentage differences)}}
    \label{tab:cliffs_energy}
    \begin{tabular}{llrrr}
        \textbf{Dataset} & \textbf{Technique} & \textbf{p-value} & \textbf{Cliff's $\delta$} & \textbf{Percentage Change} \\
        \hline
        \hline
        \multirow{8}{*}{\textbf{GSM8k}}
                         & T1AA        & \( < 0.001 \)  & 1.00       & -72.1\%           \\  
                         & T1AB        & \( < 0.001 \)  & 1.00       &-60.0\%            \\  
                         & T1B         & \(1.0 \)  & --      & 660.1\%          \\  
                         & T2          & \( < 0.001 \)  & 1.00       & -82.0\%           \\
                         & T3C         & \( < 0.001 \)  &  0.77        & -9.9\%              \\  
                         & T3D         & \(1.0 \)   & --     & 13.3\%            \\  
                         & T3E         & \(1.0 \)  & --       & 327.0\%             \\  
                         & T4         & \( 1.0 \)  & --      & 9.16\%           \\  
                         \hline         
        \multirow{8}{*}{\textbf{MMLU}}                 
                         & T1AA         & \( < 0.001\)  & 1.00       & -94.8\%         \\  
                         & T1AB         & \( < 0.001 \) & 1.00       & -93.7\%       \\  
                         & T1B          & \( 1.0 \) & --      & 450.1\%      \\  
                         & T2           & \( < 0.001 \) & 1.00       & -92.5\%            \\
                         & T3C          & \( < 0.001 \) & 1.00       & -49.7\%            \\  
                         & T3D          & \( < 0.001 \) & 1.00      & -60.8\%             \\  
                         & T3E          & \( 1.0 \) & --       & 429.1\%              \\ 
                         & T4               & --        &--    & --      \\
        \hline
        \hline           
    \end{tabular}
\end{table}

The results for \textit{Quantization} (T3) were more nuanced. 
The base model was a 4-bit quantized Phi-4 model loaded through Ollama and for comparison, we also evaluated a 4-bit quantized version of the same model obtained from Unsloth as mentioned above.
Despite both models using the same quantization scheme, potentially due to differences in architecture and core features, the measured energy consumption differed between the two. Unfortunately, we were unable to run the 8-bit precision Phi4 model on our hardware due to hardware limitations.
Therefore, we substituted it with the smaller 8-bit Phi-4-mini model. To allow a fairer baseline comparison, we used the standard Phi-4-mini model without quantization as a second smaller baseline (baseline$_s$).
Overall, T3 was not very successful in reducing energy consumption for the reasoning tasks (GSM8k).
While the approximately 10\% reduction in median energy consumption of the 2-bit quantization (T3C) is statistically significant with a Cliff's $\delta$ of 0.75, both 4-bit (T3D) and 8-bit quantization (T3E), with the latter being compared to baseline$_s$, actually increased energy consumption.
It is possible that the complexity of these reasoning tasks is not well-suited to using quantization for energy optimizations.
However, for the Q\&A tasks (MMLU), both 2-bit and 4-bit quantization showed significant and large energy reductions (Cliff's $\delta$ of 1.0), with around 50\% and 60\% respectively.
Only 8-bit quantization (T3E) led again to major increases in energy consumption compared to the smaller baseline$_s$ (around 430\%).

Lastly, \textit{Batching} (T4) was one of the most challenging methods to implement due to the hardware limitations of the Leaplab.
While we were able to use a batch size of 2 for the reasoning dataset (GSM8k), the input sequences for the Q\&A dataset (MMLU) were too long to allow batching with the available hardware.
We therefore performed the analysis only for GSM8k.
For this dataset, the technique was not effective but even increased energy consumption by approximately 9\%.
A potential reason may be that the average GPU utilization only increased slightly from 95.5\% to 96.88\%.
It is possible that the technique would have been more impactful with lower GPU utilization in the baseline or larger batch sizes, which our hardware did not allow.

\subsection{Trade-Offs With Other QAs (RQ2)}
For RQ2, we wanted to understand how the techniques impacted the important QAs accuracy and response time and which role the output token count played in these relationships.

\subsubsection{Accuracy}
\begin{figure}[h]
    \centering
    \includegraphics[width=0.9\linewidth]{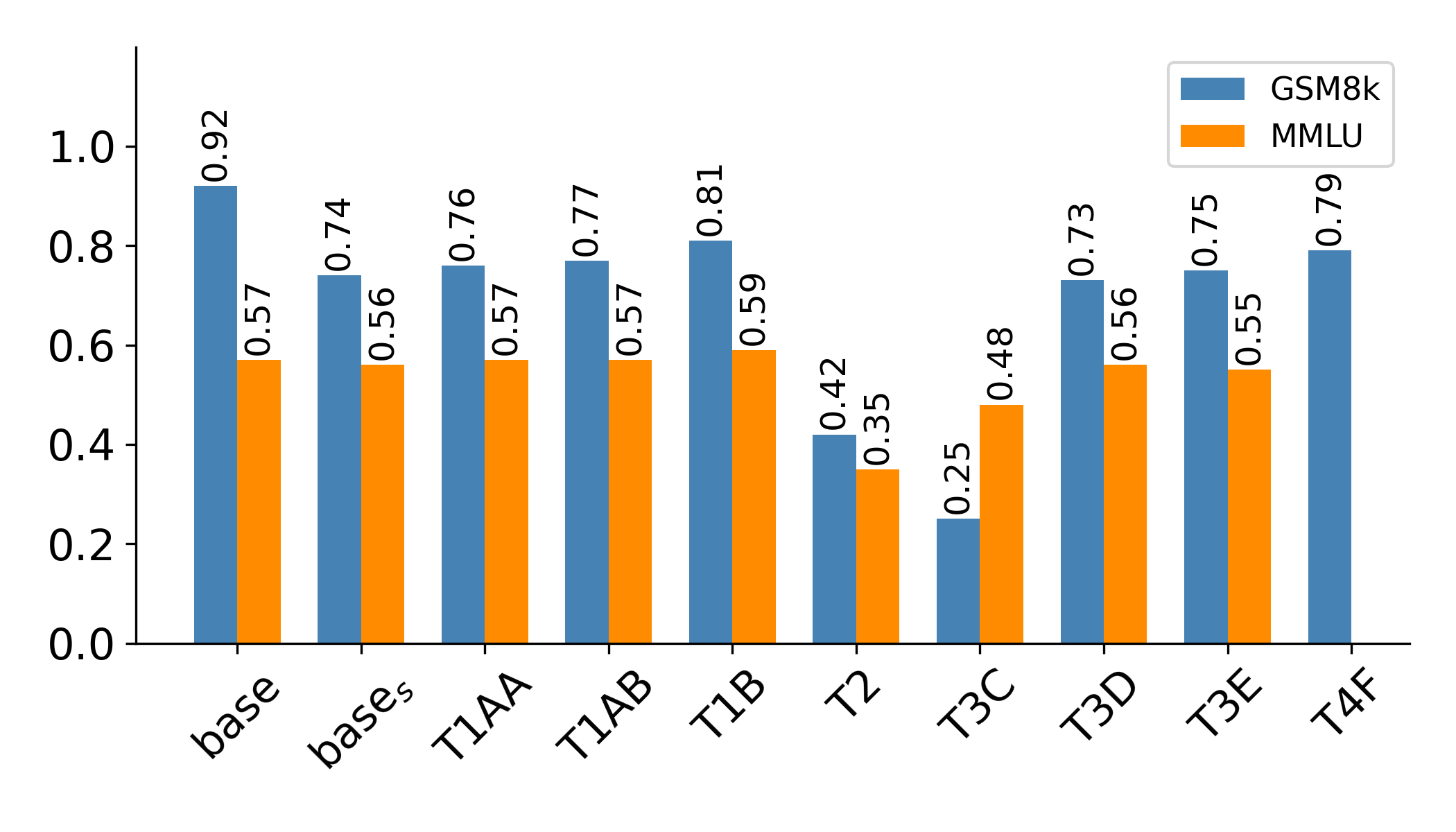}
    \caption{Accuracy on GSM8k and MMLU Datasets {\normalfont\small(base$_s$: Phi4-mini, only compared to T3E)}}
    \label{fig:accuracy}
\end{figure}

We show the accuracy per experiment configuration and dataset in Fig.~\ref{fig:accuracy}.
Overall, all techniques led to at least some accuracy drops for the GSM8k reasoning dataset (some of them substantial), while the reductions were less pronounced for the Q\&A tasks (MMLU).
Notably, PromptWizard (T2), which was one of the most effective techniques to reduce energy consumption, also led to a substantial reduction of accuracy in both datasets.
Compared to the baseline, accuracy dropped from 0.92 to 0.42 for GSM8k and from 0.57 to 0.35 for MMLU.
A similar effect is visible for 2-bit Quantization (T3C), where accuracy dropped to 0.25 (GSM8k) and 0.48 (MMLU).
Conversely, \textit{Minion} (T1B), which had by far the largest negative impact on energy consumption, achieved the best accuracy results of all techniques, namely 0.81 for GSM8k and 0.59 for MMLU, which even improved the baseline by 0.02.
However, the NPCC configurations (T1AA and T1AB), which were also among the most effective energy reduction techniques, had only slight drops in accuracy for GSM8k (0.76 and 0.77) and remarkably no accuracy reduction for MMLU (0.57), underlining their practical Green AI relevance.
Lastly, decreases in accuracy were also less pronounced for 4-bit (T3D) and 8-bit quantization (T3E) as well as batching (T4), but these techniques were also not effective in reducing energy consumption.

\subsubsection{Response Time}
\begin{figure}[h]
    \centering
    \includegraphics[width=1\linewidth]{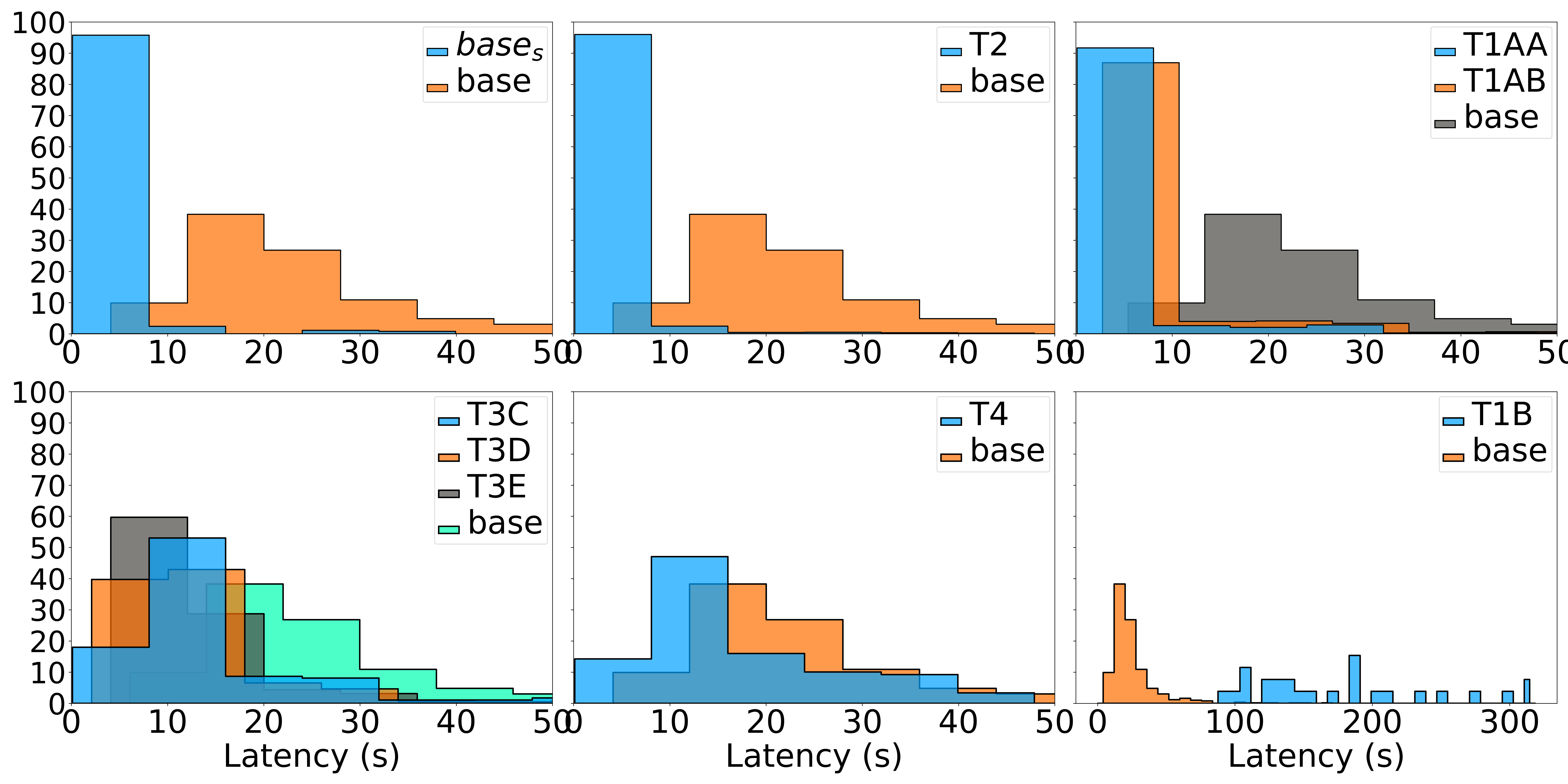}
    \caption{Response Time Distribution (s) per Experiment Configuration for Both Datasets Combined}
    \label{fig:latency-distribution}
\end{figure}

We show the distribution of response time for the various techniques compared to the baseline in Fig.~\ref{fig:latency-distribution} and the respective descriptive statistics in Table~\ref{table:descriptive_latency}.
Most of the treatments substantially reduced the response time and concentrated the distribution around lower values.
For example, PromptWizard (T2) showed a sharp shift toward near-zero response time (median of 1.28 s for GSM8k), outperforming the baseline where response times spread over a much wider range (median of 15.75 s).
Similarly, NPCC 0.3 (T1AA) and NPCC 0.6 (T1AB) showed improved response time distributions compared to the baseline.
For the three quantization techniques (T3C, T3D, T3E), we observe slightly left-shifted or almost the same distributions compared to the baseline.
However, these techniques only had lower median response times in the MMLU benchmark, not for GSM8k.
Batching (T4) managed to slightly reduce median response times compared to the baseline (from 15.75 to 13.35 s).
Finally, the Minion technique (T1B) caused a substantial response time increase, with most of its distribution falling between 100 and 300 s per request.

\begin{table}
    \centering
    \small
    \caption{Descriptive Statistics for Response Time per Prompt (s) and Experiment Configuration {\normalfont\small(baseline$_s$: Phi4-mini, only compared to T3E)}}
    \label{table:descriptive_latency}
    \begin{tabular}{llrrrrr}
        \textbf{Dataset} & \textbf{Technique} & \textbf{Mean} & \textbf{STD} & \textbf{Min} & \textbf{Median} & \textbf{Max} \\
        \hline
        \hline
        \multirow{10}{*}{\textbf{GSM8k}} 
        & T1AA & 3.25 & 2.06 & 0.42 & 2.60 & 10.63 \\  
        & T1AB & 3.81 & 3.12 & 0.42 & 2.64 & 15.00 \\  
        & T1B & 137.8 & 42.8 & 93.7 & 126 & 234.4 \\  
        & T2 & 1.62 & 0.89 & 0.99 & 1.28 & 4.46 \\  
        & T3C & 24.33 & 0.89 & 23.49 & 23.87 & 25.79 \\  
        & T3D & 25.51 & 2.16 & 20.66 & 25.74 & 28.26 \\  
        & T3E & 18.05 & 10.88 & 1.05 & 23.70 & 28.26 \\  
        & T4 & 16.43 & 8.73 & 5.18 & 13.35 & 36.88 \\  
        & baseline & 17.00 & 7.13 & 4.52 & 15.75 & 36.76 \\ 
        & baseline$_s$ & 3.07 & 1.67 & 0.12 & 2.88 & 7.33 \\
        \hline
        \multirow{8}{*}{\textbf{MMLU}} 
        & T1AA & 0.45 & 0.24 & 0.22 & 0.34 & 1.30 \\  
        & T1AB & 0.44 & 0.23 & 0.22 & 0.34 & 1.27 \\  
        & T1B & 199 & 71.4 & 97.7 & 184.6 & 318.5 \\  
        & T2 & 1.41 & 0.25 & 0.94 & 1.38 & 2.20 \\  
        & T3C & 13.23 & 6.00 & 6.26 & 11.93 & 32.93 \\  
        & T3D & 9.83 & 4.23 & 3.83 & 8.73 & 25.25 \\  
        & T3E & 7.35 & 4.58 & 0.99 & 6.49 & 23.06 \\  
        & T4 & -- & -- & -- & -- & -- \\  
        & baseline & 29.91 & 24.62 & 3.76 & 21.77 & 105 \\
        & baseline$_s$& 0.45 & 0.24 & 0.22 & 0.34 & 1.28 \\
        \hline
        \hline
    \end{tabular}
\end{table}

\subsubsection{Output Token Count}
\begin{figure}[h]
    \centering
    \includegraphics[width=0.8\linewidth]{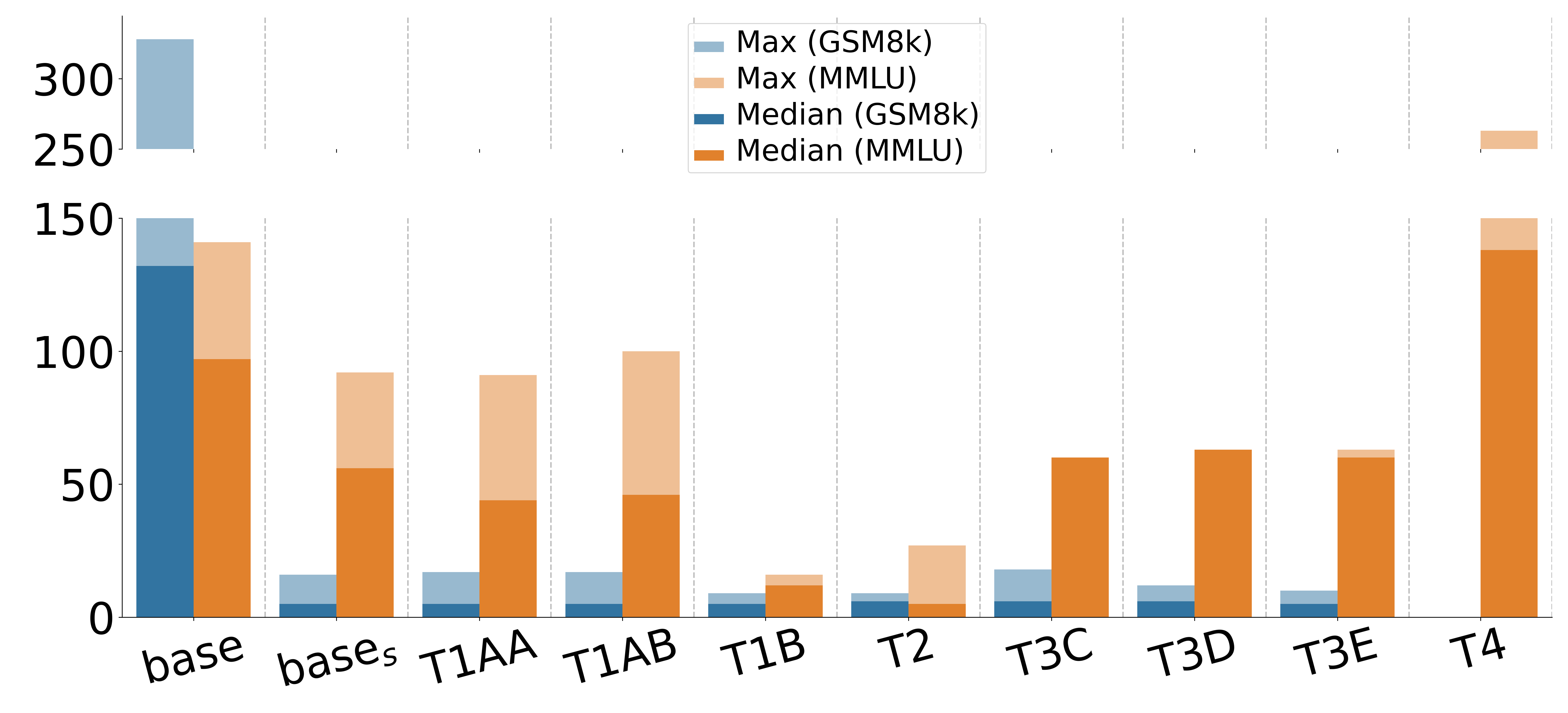}
    \caption{Max and Median Output Token Count Per Prompt}
    \label{fig:tokenall}
\end{figure}

As presented in Fig.~\ref{fig:tokenall}, median and maximum output token counts varied depending on dataset and  treatment.
Except for Batching (T4), which naturally led to more tokens per prompt by combining queries, the baseline had the highest token counts per prompt among all techniques.
While it exhibited similar median token counts for both datasets, most techniques showed a substantially lower token count for the Q\&A dataset (MMLU).
Moreover, not all techniques that reduced the token count of the final answer compared to the baseline were also effective in reducing the overall energy consumption of the system.
For example, Minion (T1B), as the most energy-consuming treatment, produced the shortest final answer.
However, during the interaction between the small and large models, a large number of tokens were generated, which we did not include in our analysis.
Nonetheless, this underlines that token count is not always a reliable proxy for energy consumption, especially not in complex distributed systems with multiple communicating LLM components.

\subsubsection{Relationships Between Dependent Variables}
\begin{figure}
    \centering
    \begin{tabular}{lll}
        \includegraphics[width=.4\linewidth]{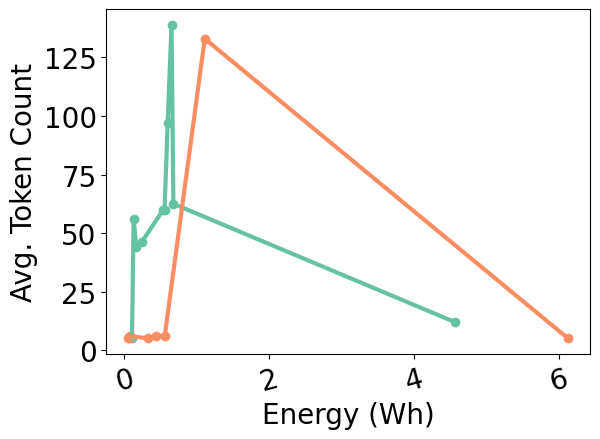} & 
        \includegraphics[width=.4\linewidth]{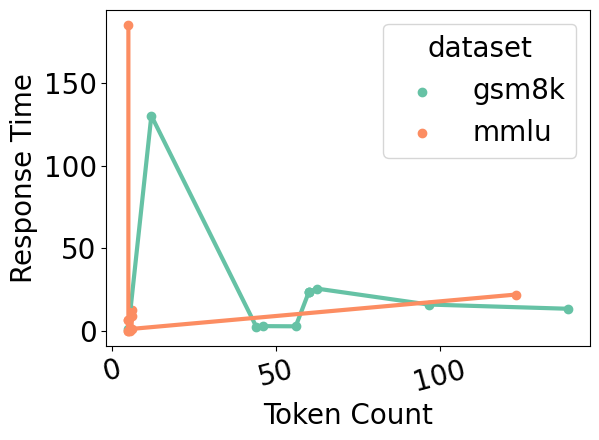} & \\
        \includegraphics[width=.4\linewidth]{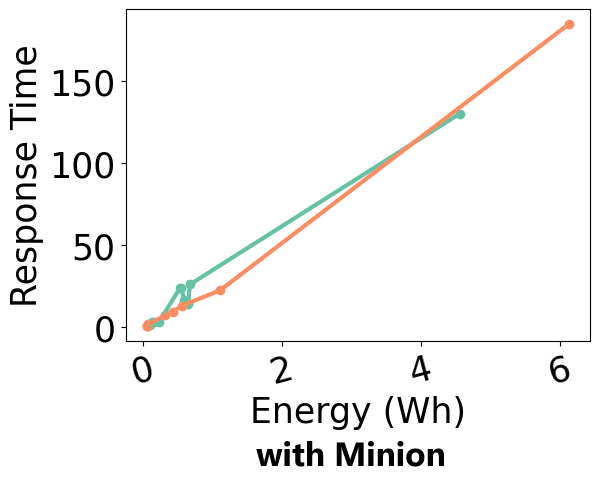} & 
        \includegraphics[width=.4\linewidth]{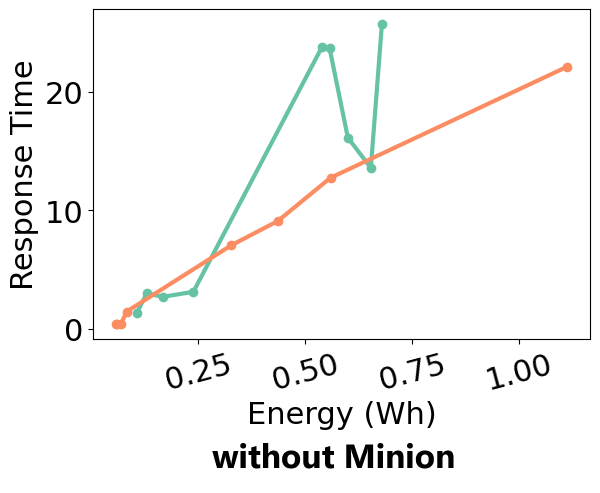}\\
    \end{tabular}
    \caption{Relationships between Output Token Count, Response Time, and Energy Consumption}
    \label{fig:others_relationship}
\end{figure}

Finally, we also analyzed the relationships between our dependent variables for the studied system variations (see Fig.~\ref{fig:others_relationship}).
Regarding \textit{response time and energy consumption} (bottom row), there first appears to be a nearly linear relationship in both datasets.
However, this relationship is heavily skewed by the Minion (T1B) results.
When we remove T1B as an outlier (bottom right), we see contrasting patterns for the two datasets.
While the Q\&A dataset (MMLU) continues to show a nearly linear trend, the reasoning dataset (GSM8k) exhibits strong fluctuations.
This shows that response time is also not always a reliable proxy for energy consumption, e.g., a technique that leads to higher median response times, like T3C for GSM8k, can still consume significantly less energy.

The relationship between \textit{response time and output token count} (top right) also exhibits distinct patterns for the two datasets.
The Q\&A dataset (MMLU) mostly shows an increase in response time as the token count rises, except for the mentioned Minion outlier.
In contrast, the graph for GSM8k is highly volatile, with significant peaks and drops in response time even as the token count remains fairly similar.
This makes it challenging to establish a clear relationship between these two quality attributes.

Lastly, the relationship between \textit{energy consumption and output token count} (top left) is similarly hard to characterize.
In both datasets, the graphs are fairly unpredictable, with several sudden peaks or drops, or energy increasing while token count stays rather stable.
This reinforces our previous observation that token count is also not a fully reliable predictor for energy reduction techniques in complex LLM-based applications.

\section{Discussion}\label{s:discussion}
In this section, we summarize the main findings and implications of our research results for practitioners and researchers.
When analyzing the results for \textbf{energy consumption}, we found that some of the selected optimizations, like the NPCC techniques, prompt optimization, or 2-bit quantization, significantly reduced energy consumption, while others led to a considerable increase. 
Specifically, we observed that the Minion (T1B) technique was associated with the highest energy consumption and response time.
This outcome is primarily due to inefficient communication between the small and large models, which requires at least two rounds of communication for each iteration.
Furthermore, we observed that in some cases, the small model generated an excessive number of tokens for tasks.
For example, an input prompt consisting of 19 tokens made the small model generate more than 10,000 tokens for the larger model, even when the temperature parameter was set to 0.
\citet{minions} mentioned similar issues as downsides of their technique, but we observed that this also substantially increased energy consumption.
Based on these findings, it is important to limit the number of tokens generated by the models and minimize the number of communication rounds between them.

When \textbf{considering accuracy and energy consumption together}, \textit{Small and Large Model Collaboration} via the NPCC techniques (T1AA, T1AB) emerged as the most energy-efficient optimization.
While T1AA and T1AB achieved significant energy reductions with large effects, they also reached the third and fourth highest accuracy scores for GSM8k and even matched the accuracy of the baseline on MMLU.
This shows that dynamically using a smaller model for appropriate queries does not have to lead to a substantial decrease in system accuracy.
We therefore recommend this technique as a practical way to improve the energy efficiency of LLM-based applications.
One open question is how to find the ideal prompt complexity threshold for specific prompt profiles.
While we included two thresholds in our study (0.3 and 0.6), conducting more fine-grained experiments in this direction or providing actionable guidance or tooling to help practitioners choose a threshold for their application is a promising opportunity for future research.

Unfortunately, the results also imply that the assumed \textbf{trade-off relationship between accuracy and energy consumption still holds in many cases}, e.g., the treatments consuming the most energy, like Minion (T1B) and the baseline, also had the highest accuracy scores.
Additionally, PromptWizard (T2) and 2-bit quantization (T3C) significantly and strongly reduced energy usage but remain of limited practical use due to their unacceptable reductions in accuracy.
While PromptWizard was effective at making prompts more efficient, key information was sometimes removed, which led to a significant decrease in accuracy.
These findings highlight that the largest energy reductions are not enough for industry usage.

Lastly, our results indicate that batching (T4) did not have a very strong effect on any dependent variable but was also not an effective technique to reduce energy consumption, as it even increased.
However, our hardware constraints only allowed a limited exploration of implementation variations of this technique.
GPU utilization as the expected benefit of batching could not increase as expected, as it was already operating above 95\% utilization for the baseline.
Accordingly, the minor increase in GPU allocation was insufficient to reduce energy consumption.
This may imply that batching effectiveness for energy optimization is highly dependent on available hardware resources and baseline utilization levels.

\smallskip

\answerbox{
  \textbf{Main Takeaways:} \newline
  \hspace*{3mm}
  \begin{minipage}{0.94\linewidth}
    - Dynamically routing queries to a smaller model based on prompt complexity can lower energy consumption without reducing accuracy for many tasks (T1A). Large models are not always necessary.\newline
    - \textbf{But:} recurring communication between small and large models and excessive token generation increases both energy consumption and response times (T1B).\newline
    - Prompt optimization (T2) trades off accuracy for energy consumption, sometimes to an unacceptable degree.\newline
    - The success and effect of optimization techniques depend on both the dataset and the used hardware.\newline
    - While reducing the energy use of LLM-based applications is easy, \textbf{optimizing energy efficiency remains difficult}, i.e., reducing energy usage without harming other QAs.
  \end{minipage}
}

\section{Threats to Validity}\label{sec:threats}
This section discusses threats to the validity of this work according to the categories introduced by \citet{Wohlin2024}.

\textit{Internal Validity}:
Limitations in hardware resources could have introduced flaws in the measurements.
In fact, during experimentation with batching configurations, we observed that memory limits were exceeded for some prompts.
We handled these cases with an exception mechanism that helped mitigate the impact of invalid trials on the results.
Additionally, we repeated each trial five times.
The execution order of trials was randomized, and a five-minute cool-down period was included between runs to dissipate accumulated energy and restore the system state before the next execution.
Moreover, our experimental setting included a hypervisor, namely VMware ESXi, which could have introduced some overhead in the energy measurement.
We mitigated the effect of such overhead on the energy values by measuring the whole machine on which ChatSBP was deployed.
Additionally, we ensured that the testbed only ran ChatSBP and no other resource-intensive processes.
While these measures should have prevented major confounding factors in the experiment environment, it is possible that minor random fluctuations could have occurred.
We are confident that this would not have changed the conclusions of the analysis.

\textit{External Validity}:
The testbed of our experiment (LeapLab) did not fully reflect the complexity of a real-world execution environment found in large-scale industry deployments.
One of the main limitations can be found in the hardware resources, which could not run large-scale AI workloads.
Additionally, our study was executed for a single industry case, ChatSBP, which does not represent the full landscape of AI-driven industry tools.
However, our investigation is meant to show the impact of empirical software engineering in practice and to describe how it can be applied in an industrial setting.
Likewise, the selection of optimization techniques was constrained by their suitability for ChatSBP and their relevance to the specific problems faced by its employees.
Consequently, we cannot fully guarantee their applicability or relevance in other contexts.
Nevertheless, the addressed problem is recurrent in modern AI-driven applications, where it is common to pre-process data before sending it to an AI module deployed in the cloud.
Lastly, although the tasks included in the chosen datasets may not cover the full range of complexities found in real-life industrial applications, they were selected because they capture representative scenarios that allow us to investigate whether the evaluated techniques have the potential to improve the energy efficiency of ChatSBP.

\textit{Conclusion Validity}:
We selected appropriate statistical methods for non-normally distributed data.
We also applied the Holm-Bonferroni correction~\cite{Holm-Bonferroni} to minimize Type I errors due to multiple comparisons.
This further solidified the validity of our results.

\section{Conclusion}
In this study, we systematically evaluated the effectiveness of various energy optimization techniques for an industrial chatbot application.
By implementing and analyzing techniques such as \textit{Small and Large Model Collaboration}, \textit{Prompt Optimization}, \textit{Quantization}, and \textit{Batching}, we demonstrated that the energy consumption of LLM-based applications can be significantly reduced with most techniques.
However, we also highlighted the challenge of balancing the trade-offs between energy consumption and other QAs, especially accuracy.
While several effective techniques like \textit{Prompt Optimization} reduced accuracy to unacceptable levels, \textit{Small and Large Model Collaboration} via NPCC achieved substantial reductions in energy usage while maintaining acceptable accuracy scores and response time, making it a prime candidate for industry adoption.

Our findings also highlighted limitations in both the technical environment and the generalizability of results.
The complexities of industrial deployment, including hardware constraints, diverse workload patterns, or even different models, may influence the impact of any energy optimization technique.
Thus, evaluation of optimization techniques across a broader range of LLM architectures, application domains, and production environments could provide further insights and guidance and should be the target of future work.
Many proposed techniques, like optimizing energy-hungry prompt keywords~\cite{adamska2025greenprompting} or using generation directives~\cite{li_sprout_2024}, seem promising and require additional research to understand potential trade-offs better.
Additionally, novel approaches, such as continuous batching with more suitable hardware resources, could also provide further insights regarding the impact of batching on energy efficiency.
Furthermore, an investigation into user experience metrics with absolute real-world settings would also contribute to the practical adoption of these energy optimization techniques.
By carefully balancing energy consumption with other QAs, the field can move closer to being more energy-efficient.
To support future studies and replications, we make our artifacts available online.\textsuperscript{\ref{fn:replication-pckg}}

\begin{acks}
We kindly thank Cornelis Mol for his support in improving the visualizations of this study, which played an important role in making the presentation more effective.
\end{acks}

\bibliographystyle{ACM-Reference-Format}
\bibliography{references}

@book{Wohlin2024,
  author    = {Claes Wohlin and Per Runeson and Martin Höst and Magnus C. Ohlsson and Björn Regnell and Anders Wesslén},
  title     = {Experimentation in Software Engineering},
  edition   = {2nd},
  publisher = {Springer Berlin Heidelberg},
  address   = {Berlin, Heidelberg},
  year      = {2024},
  url       = {https://link.springer.com/10.1007/978-3-662-69306-3},
  doi       = {10.1007/978-3-662-69306-3},
  pages = "109--111"

}

@misc{adamska2025greenprompting,
      title={Green Prompting}, 
      author={Marta Adamska and Daria Smirnova and Hamid Nasiri and Zhengxin Yu and Peter Garraghan},
      year={2025},
      eprint={2503.10666},
      archivePrefix={arXiv},
      primaryClass={cs.CL},
      url={https://arxiv.org/abs/2503.10666}, 
}

@INPROCEEDINGS{Geens2024,
  author={Geens, Robin and Shi, Man and Symons, Arne and Fang, Chao and Verhelst, Marian},
  booktitle={2024 IEEE 37th International System-on-Chip Conference (SOCC)}, 
  title={Energy Cost Modelling for Optimizing Large Language Model Inference on Hardware Accelerators}, 
  year={2024},
  volume={},
  number={},
  pages={1-6},

  doi={10.1109/SOCC62300.2024.10737844}
}

@misc{poddar2025measure_nlp,
      title={Towards Sustainable NLP: Insights from Benchmarking Inference Energy in Large Language Models}, 
      author={Soham Poddar and Paramita Koley and Janardan Misra and Sanjay Podder and Niloy Ganguly and Saptarshi Ghosh},
      year={2025},
      eprint={2502.05610},
      archivePrefix={arXiv},
      primaryClass={cs.CL},
      url={https://arxiv.org/abs/2502.05610}, 
}

@misc{minions,
      title={Minions: Cost-efficient Collaboration Between On-device and Cloud Language Models}, 
      author={Avanika Narayan and Dan Biderman and Sabri Eyuboglu and Avner May and Scott Linderman and James Zou and Christopher Re},
      year={2025},
      eprint={2502.15964},
      archivePrefix={arXiv},
      primaryClass={cs.LG},
      url={https://arxiv.org/abs/2502.15964}, 
}

@misc{agarwal2024promptwizard,
      title={PromptWizard: Task-Aware Prompt Optimization Framework}, 
      author={Eshaan Agarwal and Joykirat Singh and Vivek Dani and Raghav Magazine and Tanuja Ganu and Akshay Nambi},
      year={2024},
      eprint={2405.18369},
      archivePrefix={arXiv},
      primaryClass={cs.CL},
      url={https://arxiv.org/abs/2405.18369}, 
}

@misc{ma2025onebit,
      title={BitNet b1.58 2B4T Technical Report}, 
      author={Shuming Ma and Hongyu Wang and Shaohan Huang and Xingxing Zhang and Ying Hu and Ting Song and Yan Xia and Furu Wei},
      year={2025},
      eprint={2504.12285},
      archivePrefix={arXiv},
      primaryClass={cs.CL},
      url={https://arxiv.org/abs/2504.12285}, 
}

@ARTICLE{argerichQuantiz,
  author={Argerich, Mauricio Fadel and Patiño-Martínez, Marta},
  journal={IEEE Access}, 
  title={Measuring and Improving the Energy Efficiency of Large Language Models Inference}, 
  year={2024},
  volume={12},
  number={},
  pages={80194-80207},
  doi={10.1109/ACCESS.2024.3409745}
}

@misc{ma2024era1bitllmslarge,
      title={The Era of 1-bit LLMs: All Large Language Models are in 1.58 Bits}, 
      author={Shuming Ma and Hongyu Wang and Lingxiao Ma and Lei Wang and Wenhui Wang and Shaohan Huang and Li Dong and Ruiping Wang and Jilong Xue and Furu Wei},
      year={2024},
      eprint={2402.17764},
      archivePrefix={arXiv},
      primaryClass={cs.CL},
      url={https://arxiv.org/abs/2402.17764}, 
}

@inproceedings{mmlu,
  title={Measuring Massive Multitask Language Understanding},
  author={Hendrycks, Dan and Burns, Collin and Basart, Steven and Zou, Andy and Mazeika, Mantas and Song, Dawn and Steinhardt, Jacob},
  booktitle={International Conference on Learning Representations}
}

@misc{gsm8k,
      title={Training Verifiers to Solve Math Word Problems}, 
      author={Karl Cobbe and Vineet Kosaraju and Mohammad Bavarian and Mark Chen and Heewoo Jun and Lukasz Kaiser and Matthias Plappert and Jerry Tworek and Jacob Hilton and Reiichiro Nakano and Christopher Hesse and John Schulman},
      year={2021},
      eprint={2110.14168},
      archivePrefix={arXiv},
      primaryClass={cs.LG},
      url={https://arxiv.org/abs/2110.14168}, 
}

@misc{microsoft2025phi4minitechnicalrepo,
      title={Phi-4-Mini Technical Report: Compact yet Powerful Multimodal Language Models via Mixture-of-LoRAs}, 
      author={Microsoft and : and Abdelrahman Abouelenin and Atabak Ashfaq and Adam Atkinson and Hany Awadalla and Nguyen Bach and Jianmin Bao and Alon Benhaim and Martin Cai and Vishrav Chaudhary and Congcong Chen and Dong Chen and Dongdong Chen and Junkun Chen and Weizhu Chen and Yen-Chun Chen and Yi-ling Chen and Qi Dai and Xiyang Dai and Ruchao Fan and Mei Gao and Min Gao and Amit Garg and Abhishek Goswami and Junheng Hao and Amr Hendy and Yuxuan Hu and Xin Jin and Mahmoud Khademi and Dongwoo Kim and Young Jin Kim and Gina Lee and Jinyu Li and Yunsheng Li and Chen Liang and Xihui Lin and Zeqi Lin and Mengchen Liu and Yang Liu and Gilsinia Lopez and Chong Luo and Piyush Madan and Vadim Mazalov and Arindam Mitra and Ali Mousavi and Anh Nguyen and Jing Pan and Daniel Perez-Becker and Jacob Platin and Thomas Portet and Kai Qiu and Bo Ren and Liliang Ren and Sambuddha Roy and Ning Shang and Yelong Shen and Saksham Singhal and Subhojit Som and Xia Song and Tetyana Sych and Praneetha Vaddamanu and Shuohang Wang and Yiming Wang and Zhenghao Wang and Haibin Wu and Haoran Xu and Weijian Xu and Yifan Yang and Ziyi Yang and Donghan Yu and Ishmam Zabir and Jianwen Zhang and Li Lyna Zhang and Yunan Zhang and Xiren Zhou},
      year={2025},
      eprint={2503.01743},
      archivePrefix={arXiv},
      primaryClass={cs.CL},
      url={https://arxiv.org/abs/2503.01743}, 
}

@misc{abdin2024phi4technicalreport,
      title={Phi-4 Technical Report}, 
      author={Marah Abdin and Jyoti Aneja and Harkirat Behl and Sébastien Bubeck and Ronen Eldan and Suriya Gunasekar and Michael Harrison and Russell J. Hewett and Mojan Javaheripi and Piero Kauffmann and James R. Lee and Yin Tat Lee and Yuanzhi Li and Weishung Liu and Caio C. T. Mendes and Anh Nguyen and Eric Price and Gustavo de Rosa and Olli Saarikivi and Adil Salim and Shital Shah and Xin Wang and Rachel Ward and Yue Wu and Dingli Yu and Cyril Zhang and Yi Zhang},
      year={2024},
      eprint={2412.08905},
      archivePrefix={arXiv},
      primaryClass={cs.CL},
      url={https://arxiv.org/abs/2412.08905}, 
}

@INPROCEEDINGS{debated_batching_image,
  author={Yarally, Tim and Cruz, Luís and Feitosa, Daniel and Sallou, June and van Deursen, Arie},
  booktitle={2023 49th Euromicro Conference on Software Engineering and Advanced Applications (SEAA)}, 
  title={Batching for Green AI - An Exploratory Study on Inference}, 
  year={2023},
  volume={},
  number={},
  pages={112-119},
  doi={10.1109/SEAA60479.2023.00026}}

@inproceedings{renze-2024-effect,
    title = "The Effect of Sampling Temperature on Problem Solving in Large Language Models",
    author = "Renze, Matthew",
    editor = "Al-Onaizan, Yaser  and
      Bansal, Mohit  and
      Chen, Yun-Nung",
    booktitle = "Findings of the Association for Computational Linguistics: EMNLP 2024",
    month = nov,
    year = "2024",
    address = "Miami, Florida, USA",
    publisher = "Association for Computational Linguistics",
    url = "https://aclanthology.org/2024.findings-emnlp.432/",
    doi = "10.18653/v1/2024.findings-emnlp.432",
    pages = "7346--7356"
}

@inproceedings{vinutha2018iqr,
  author    = {H. Vinutha and B. Poornima and B. Sagar},
  title     = {Detection of outliers using interquartile range technique from intrusion dataset},
  booktitle = {Information and Decision Sciences: Proceedings of the 6th International Conference on FICTA},
  year      = {2018},
  publisher = {Springer},
  pages     = {511--518}
}

@article{shapiro,
 ISSN = {00063444, 14643510},
 URL = {http://www.jstor.org/stable/2333709},
 author = {S. S. Shapiro and M. B. Wilk},
 journal = {Biometrika},
 number = {3/4},
 pages = {591--611},
 publisher = {[Oxford University Press, Biometrika Trust]},
 title = {An Analysis of Variance Test for Normality (Complete Samples)},
 urldate = {2025-06-19},
 volume = {52},
 year = {1965}
}

@article{q-qtest,
  title={Variations of Q--Q plots: The power of our eyes!},
  author={Loy, Adam and Follett, Lendie and Hofmann, Heike},
  journal={The American Statistician},
  volume={70},
  number={2},
  pages={202--214},
  year={2016},
  publisher={Taylor \& Francis}
}

@article{Mann-Whitney,
  title={Mann-whitney U test},
  author={McKnight, Patrick E and Najab, Julius},
  journal={The Corsini encyclopedia of psychology},
  pages={1--1},
  year={2010},
  publisher={Wiley Online Library}
}

@article{Holm-Bonferroni,
  title={Bonferroni correction},
  author={Weisstein, Eric W},
  journal={https://mathworld. wolfram. com/},
  year={2004},
  publisher={Wolfram Research, Inc.}
}

@article{wang2024comprehensivesurveysmalllanguage,
  title={A comprehensive survey of small language models in the era of large language models: Techniques, enhancements, applications, collaboration with llms, and trustworthiness},
  author={Wang, Fali and Zhang, Zhiwei and Zhang, Xianren and Wu, Zongyu and Mo, Tzuhao and Lu, Qiuhao and Wang, Wanjing and Li, Rui and Xu, Junjie and Tang, Xianfeng and others},
  journal={ACM Transactions on Intelligent Systems and Technology},
  year={2024},
  publisher={ACM New York, NY}
}

@inproceedings{Chien2023,
author = {Chien, Andrew A and Lin, Liuzixuan and Nguyen, Hai and Rao, Varsha and Sharma, Tristan and Wijayawardana, Rajini},
title = {Reducing the Carbon Impact of Generative AI Inference (today and in 2035)},
year = {2023},
isbn = {9798400702426},
publisher = {Association for Computing Machinery},
address = {New York, NY, USA},
url = {https://doi.org/10.1145/3604930.3605705},
doi = {10.1145/3604930.3605705},
booktitle = {Proceedings of the 2nd Workshop on Sustainable Computer Systems},
articleno = {11},
numpages = {7},
location = {Boston, MA, USA},
series = {HotCarbon '23}
}

@inproceedings{NEURIPS2023_0df38cd1,
 author = {Chee, Jerry and Cai, Yaohui and Kuleshov, Volodymyr and De Sa, Christopher M},
 booktitle = {Advances in Neural Information Processing Systems},
 editor = {A. Oh and T. Naumann and A. Globerson and K. Saenko and M. Hardt and S. Levine},
 pages = {4396--4429},
 publisher = {Curran Associates, Inc.},
 title = {QuIP: 2-Bit Quantization of Large Language Models With Guarantees},
 url = {https://proceedings.neurips.cc/paper_files/paper/2023/file/0df38cd13520747e1e64e5b123a78ef8-Paper-Conference.pdf},
 volume = {36},
 year = {2023}
}

@misc{nvidia_prompt_task_complexity_classifier,
  title        = {Prompt Task and Complexity Classifier},
  author       = {{NVIDIA NeMo Curator Team}},
  howpublished = {\url{https://huggingface.co/nvidia/prompt-task-and-complexity-classifier}},
  note         = {Accessed: 2025-06-28},
}

@misc{azure-openai-pricing,
  author       = {{Microsoft Azure}},
  title        = {Pricing – Azure OpenAI Service},
  year         = 2025,
  url          = {https://azure.microsoft.com/en-us/pricing/details/cognitive-services/openai-service},
  note         = {Accessed: 2025-06-30}
}

@misc{azure-phi3-pricing,
  author       = {{Microsoft Azure}},
  title        = {Pricing – Phi-3 Models},
  year         = 2025,
  url          = {https://azure.microsoft.com/en-us/pricing/details/phi-3/},
  note         = {Accessed: 2025-06-30}
}

@misc{llm_security_cool_paper,
      title={Mapping LLM Security Landscapes: A Comprehensive Stakeholder Risk Assessment Proposal}, 
      author={Rahul Pankajakshan and Sumitra Biswal and Yuvaraj Govindarajulu and Gilad Gressel},
      year={2024},
      eprint={2403.13309},
      archivePrefix={arXiv},
      primaryClass={cs.CR},
      url={https://arxiv.org/abs/2403.13309}, 
}

@article{chang_survey_2024,
    title = {A {Survey} on {Evaluation} of {Large} {Language} {Models}},
    volume = {15},
    issn = {2157-6904, 2157-6912},
    url = {https://dl.acm.org/doi/10.1145/3641289},
    doi = {10.1145/3641289},
    language = {en},
    number = {3},
    urldate = {2024-11-17},
    journal = {ACM Transactions on Intelligent Systems and Technology},
    author = {Chang, Yupeng and Wang, Xu and Wang, Jindong and Wu, Yuan and Yang, Linyi and Zhu, Kaijie and Chen, Hao and Yi, Xiaoyuan and Wang, Cunxiang and Wang, Yidong and Ye, Wei and Zhang, Yue and Chang, Yi and Yu, Philip S. and Yang, Qiang and Xie, Xing},
    month = jun,
    year = {2024},
    pages = {1--45},
}

@inproceedings{jarvenpaa_synthesis_2024,
    address = {Lisbon Portugal},
    title = {A {Synthesis} of {Green} {Architectural} {Tactics} for {ML}-{Enabled} {Systems}},
    copyright = {All rights reserved},
    isbn = {9798400704994},
    url = {https://dl.acm.org/doi/10.1145/3639475.3640111},
    doi = {10.1145/3639475.3640111},
    language = {en},
    urldate = {2024-06-08},
    booktitle = {Proceedings of the 46th {International} {Conference} on {Software} {Engineering}: {Software} {Engineering} in {Society}},
    publisher = {ACM},
    author = {Järvenpää, Heli and Lago, Patricia and Bogner, Justus and Lewis, Grace and Muccini, Henry and Ozkaya, Ipek},
    month = apr,
    year = {2024},
    pages = {130--141},
}

@inproceedings{stojkovic2025dynamollm,
  title={Dynamollm: Designing llm inference clusters for performance and energy efficiency},
  author={Stojkovic, Jovan and Zhang, Chaojie and Goiri, {\'I}{\~n}igo and Torrellas, Josep and Choukse, Esha},
  booktitle={2025 IEEE International Symposium on High Performance Computer Architecture (HPCA)},
  pages={1348--1362},
  year={2025},
  organization={IEEE}
}

@INPROCEEDINGS{Khan2025,
  author={Khan, Tahniat and Motie, Soroor and Kocak, Sedef Akinli and Raza, Shaina},
  booktitle={2025 IEEE Conference on Artificial Intelligence (CAI)}, 
  title={Optimizing Large Language Models: Metrics, Energy Efficiency, and Case Study Insights}, 
  year={2025},
  volume={},
  number={},
  pages={370-375},
  doi={10.1109/CAI64502.2025.00067}}

@INPROCEEDINGS{Rubei2025,
  author={Rubei, Riccardo and Moussaid, Aicha and Di Sipio, Claudio and Di Ruscio, Davide},
  booktitle={2025 IEEE/ACM 9th International Workshop on Green and Sustainable Software (GREENS)}, 
  title={Prompt engineering and its implications on the energy consumption of Large Language Models}, 
  year={2025},
  volume={},
  number={},
  pages={60-67},
  doi={10.1109/GREENS66463.2025.00014}}

@INPROCEEDINGS{Kaushik2025,
  author={Kaushik, Bhanu and Taneja, Aman and Dahiya, Sonika},
  booktitle={2025 8th International Conference on Computing Methodologies and Communication (ICCMC)}, 
  title={Toward Sustainable AI: A Review of Energy-Efficient Large Language Models}, 
  year={2025},
  volume={},
  number={},
  pages={944-951},
  doi={10.1109/ICCMC65190.2025.11140923}}

@article{Schwartz2020,
title={Green AI}, 
volume={63}, 
url={https://doi.org/10.1145/3381831}, 
DOI={10.1145/3381831}, 
number={12}, 
journal={Communications of the ACM}, 
author={Schwartz, Roy and Dodge, Jesse and Smith, Noah A. and Etzioni, Oren}, 
year={2020},
month=nov, 
pages={54–63} }

@InProceedings{Walkowiak2025,
author="Walkowiak, Tomasz",
title="Energy Efficiency in Large Language Models: An Empirical Study",
booktitle="Advances in Dependable Systems and Networks",
DOI= {10.1007/978-3-031-92734-8_22},
year="2025",
publisher="Springer Nature Switzerland",
address="Cham",
pages="221--228",
isbn="978-3-031-92734-8"
}

@InProceedings{Ggaliwango2024,
author={Marvin, Ggaliwango
and Hellen, Nakayiza
and Jjingo, Daudi
and Nakatumba-Nabende, Joyce},
title="Prompt Engineering in Large Language Models",
booktitle="Data Intelligence and Cognitive Informatics",
DOI={10.1007/978-981-99-7962-2_30},
year="2024",
publisher="Springer Nature Singapore",
address="Singapore",
pages="387--402",
isbn="978-981-99-7962-2"
}

@article{Husom2025,
author = {Husom, Erik Johannes and Goknil, Arda and Astekin, Merve and Shar, Lwin Khin and K\r{a}sen, Andre and Sen, Sagar and Mithassel, Benedikt Andreas and Soylu, Ahmet},
title = {Sustainable LLM Inference for Edge AI: Evaluating Quantized LLMs for Energy Efficiency, Output Accuracy, and Inference Latency},
year = {2025},
publisher = {Association for Computing Machinery},
address = {New York, NY, USA},
url = {https://doi.org/10.1145/3767742},
doi = {10.1145/3767742},
note = {Just Accepted},
journal = {ACM Trans. Internet Things},
month = sep
}

@article{runeson_guidelines_2009,
    title = {Guidelines for conducting and reporting case study research in software engineering},
    volume = {14},
    issn = {1382-3256},
    url = {http://link.springer.com/10.1007/s10664-008-9102-8},
    doi = {10.1007/s10664-008-9102-8},
    number = {2},
    journal = {Empirical Software Engineering},
    author = {Runeson, Per and Höst, Martin},
    month = apr,
    year = {2009},
    pmid = {28843849},
    note = {ISBN: 1382325615737616
\_eprint: 9809069v1},
    pages = {131--164},
}

@article{apsan2025generating,
  title={Generating Energy-Efficient Code via Large-Language Models--Where are we now?},
  author={Apsan, Radu and Stoico, Vincenzo and Albonico, Michel and Dhar, Rudra and Vaidhyanathan, Karthik and Malavolta, Ivano},
  journal={arXiv preprint arXiv:2509.10099},
  year={2025}
}

@article{kavathekar2025small,
  title={Small models, big tasks: An exploratory empirical study on small language models for function calling},
  author={Kavathekar, Ishan and Donakanti, Raghav and Kumaraguru, Ponnurangam and Vaidhyanathan, Karthik},
  journal={arXiv preprint arXiv:2504.19277},
  year={2025}
}

@article{duran2024energy,
  title={Energy consumption of code small language models serving with runtime engines and execution providers},
  author={Dur{\'a}n, Francisco and Martinez, Matias and Lago, Patricia and Mart{\'\i}nez-Fern{\'a}ndez, Silverio},
  journal={arXiv preprint arXiv:2412.15441},
  year={2024}
}

@article{hou_large_2024,
    title = {Large {Language} {Models} for {Software} {Engineering}: {A} {Systematic} {Literature} {Review}},
    issn = {1049-331X, 1557-7392},
    shorttitle = {Large {Language} {Models} for {Software} {Engineering}},
    url = {https://dl.acm.org/doi/10.1145/3695988},
    doi = {10.1145/3695988},
    language = {en},
    urldate = {2024-09-23},
    journal = {ACM Transactions on Software Engineering and Methodology},
    author = {Hou, Xinyi and Zhao, Yanjie and Liu, Yue and Yang, Zhou and Wang, Kailong and Li, Li and Luo, Xiapu and Lo, David and Grundy, John and Wang, Haoyu},
    month = sep,
    year = {2024},
    pages = {3695988},
}

@inproceedings{li_sprout_2024,
    address = {Miami, Florida, USA},
    title = {Sprout: {Green} {Generative} {AI} with {Carbon}-{Efficient} {LLM} {Inference}},
    shorttitle = {Sprout},
    url = {https://aclanthology.org/2024.emnlp-main.1215},
    doi = {10.18653/v1/2024.emnlp-main.1215},
    language = {en},
    urldate = {2025-12-11},
    booktitle = {Proceedings of the 2024 {Conference} on {Empirical} {Methods} in {Natural} {Language} {Processing}},
    publisher = {Association for Computational Linguistics},
    author = {Li, Baolin and Jiang, Yankai and Gadepally, Vijay and Tiwari, Devesh},
    year = {2024},
    pages = {21799--21813},
}

\end{document}